\documentstyle[preprint,aps,psfig]{revtex}

\newcommand{\be}{\begin{equation}}
\newcommand{\ee}{\end{equation}}

\tightenlines

\begin{document}

\draft
% \twocolumn[\hsize\textwidth\columnwidth\hsize\csname
% @twocolumnfalse\endcsname

\title{\bf Dynamical Transition from Triplets to Spinon Excitations:\\
A Series Expansion Study of the $J_1-J_2-\delta$ spin-half chain}
\author{Rajiv R. P. Singh}
\address{Department of Physics, University of California,
Davis, CA 95616}
\author{Zheng Weihong}
\address{School of Physics,
The University of New South Wales,
Sydney, NSW 2052, Australia.}

\date{Dec. 21, 1998}
%\date{\today}

\maketitle

\begin{abstract}
We study the spin-half Heisenberg chain with alternating nearest neighbor
interactions $J_1(1+\delta)$ and $J_1(1-\delta)$ and
a uniform second neighbor interaction $J_2=y (1-\delta)$ by series expansions
around the limit of decoupled dimers ($\delta=1$). 
By extrapolating to $\delta=0$ and tuning $y$, %ratio $J_2/J_1$, 
we study the
critical point separating the power-law and spontaneously dimerized
phases of the spin-half antiferromagnet. We then focus on the disorder
line $y=0.5$, $0\le \delta\le 1$, where the ground states
are known exactly. We calculate the triplet excitation spectrum, their spectral
weights and wavevector dependent static susceptibility along this line.
It is well known that as $\delta \to 0$, the spin-gap is still non-zero
but the triplets are replaced by spinons as the elementary
excitations. We study this dynamical transition by analyzing the
series for the spectral weight and the static susceptibility. In
particular, we show that the spectral weight for the triplets
vanishes and the static spin-susceptibility changes from a simple pole at
imaginary wavevectors to a branch cut at the transition.
\end{abstract}

% \phantom{.}
% ]

\narrowtext
\section{Introduction}
The study of quantum-disordered ground states of low dimensional
spin systems, with an absence of long-range magnetic order and a gap to
spin excitations, has attracted considerable
interest recently \cite{chn}. A question of fundamental importance
in the field is the nature of
elementary excitations in these phases.
These spin excitations could be related to simple spin-flips,
in which case they should carry spin-one, or they could represent
free spin-half excitations in an otherwise spinless background. 
The existence of such spin-half excitations
or spinons, in $d>1$ lattice models remains an outstanding open question
\cite{spinon1,spinon2}. Thus
it is important to develop suitable numerical schemes that can
look for such spin-half excitations. The purpose of this paper
is to test such a series expansion based method on a 1-dimensional model,
where the existence of such excitations is well known,
and to study the properties of the transition.

The $J_1-J_2-\delta$ spin-half chain is given by the Hamiltonian:
\be
{\cal H}=J_1\sum_i (1+(-1)^i\delta)\vec S_i\cdot\vec S_{i+1}
+J_2\sum_i \vec S_i\vec S_{i+2} \label{H}
\ee
This model has been a subject of many theoretical 
studies \cite{majumdar,shastry,J1-J2-delta}. In
particular for $\delta=0$, it is well known that the model
undergoes a phase transition from a critical phase at small
$J_2/J_1$ to a spontaneously 
dimerized phase at large $J_2/J_1$.
The critical value of $J_2/J_1$ has been accurately computed to
be $(J_2/J_1)_c =0.2411$\cite{oka92-castilla}.
It is also well known that for the nearest neighbor spin-half
chain, the presence of marginal operators lead to 
logarithmic corrections to various correlation
functions \cite{aff89}. It has been argued that these marginal operators
are absent at the transition to the dimerized phase \cite{fieldt} and in this
case the logarithmic corrections should also go away. This result
has been confirmed in previous numerical studies \cite{eggert}. Here, we present
results from series analysis, which lend further support to it.

The primary focus of this paper is on the excitation spectra
along a special line in the parameter space, where the
ground state is known exactly.
Along this line $0\le 2 J_2/J_1 =1-\delta<1$, this model has a unique
ground state consisting of singlet pairs between spins at
$S_{2i}$ and $S_{2i+1}$ \cite{shastry}. 
Having the simple ground states with no quantum fluctuations
allows us to focus on the elementary excitations of the system.
In the disconnected dimer limit, $J_2=J_1(1-\delta)/2=0$,
the elementary excitations are localized triplets, where one
of the spin-pair ($S_{2i},S_{2i+1}$) is excited to a triplet.
These triplet excitations develop a dispersion for $\delta \ne 1$,
but remain well defined for all $\delta>0$.
At $\delta=0$, the Hamiltonian has full translational symmetry,
and from the well known results of Majumdar and Ghosh
\cite{majumdar}, this symmetry is spontaneously broken leading to
two degenerate ground states. One of these two ground states is
the same as the ground state for $\delta>0$. In this case, the elementary
excitations are spinons or domain walls with spin-half, which
interpolate between the two ground states. This result was first
established through the variational calculations of Shastry and
Sutherland \cite{shastry} and has since been confirmed by 
many authors \cite{J1-J2-delta}. 

A popular way to study the spinons to triplet transition occuring
at small $\delta$, is through the binding of spinon pairs due
to the confining linear potential \cite{spinon-binding}. 
Here we will consider the
opposite point of view and study the break up of the
triplets into spinon pairs as $\delta$ goes to zero.
Thus, this method allows one to look for spinon excitations,
in models where there existence is not yet established, starting from
a limit where only triplet excitations exist. Recently such a
search for spinons was carried out in the bilayer triangular-lattice
Heisenberg model, where they were not found to be present\cite{triangle}.

Here we calculate the dispersion for the triplet excitations,
their spectral weights and the wavevector dependent static susceptibility by 
series expansions around the disconnected dimer limit. 
The series analysis clearly confirms that as $\delta\to 0$,
the spectral weights for the triplets vanishes (except very
near the dispersion maximum at $k=\pi/2$, where in the spinon
picture triplet bound states are known to exist). Furthermore,
for $\delta\ne 0$, the static susceptibility has a simple pole at
imaginary wavevectors, reflecting the quasiparticle nature of
the triplet excitations. As $\delta\to 0$, this turns into 
a branch cut reflecting the absence of triplet quasiparticles.
This study shows that the series expansion method is well suited
to studying this transition. Since this method can easily be
applied in higher dimensional systems, it gives us hope that
it can be used to search for such spin-half excitations in those
cases as well.

\section{Series Calculations}
To construct a series expansion around the limit of disconnected
dimers in powers of 
$$\lambda={1-\delta\over 1+\delta},$$ 
one can rewrite the  Hamiltonian in
Eq. (\ref{H}) in the following form:
\be
{\cal H}/(1+\delta)J_1 = H_0 + \lambda V,
\ee
where the unperturbed Hamiltonian $H_0$ and the perturbation $V$ are
\be
H_0 = \sum_i \vec S_{2i}\cdot\vec S_{2i+1},
\ee
\be
V = \sum_i \vec S_{2i-1}\cdot\vec S_{2i} + y \sum_i \vec S_i\cdot \vec S_{i+2}.
\ee
$y$ is related to $J_2$ by the relation
\be
y=J_2/(1-\delta)J_1.
\ee
The  expansions are developed  for fixed values of $y$.
The expansion methods for the wavevector dependent susceptibility \cite{he90,gel90},
the triplet dispersion \cite{gelfand}, and the spectral weight \cite{sweight}
are discussed in the literature. 
We will concentrate on the expansions for the following three 
different values of $y$:

(1) $y=$0, that is, without the second neighbor interaction;

(2) $y=y_c\equiv (J_2/J_1)_c=0.2411$, that is, the system is at the
critical point between gapped and gapless 
phases when $\lambda=1$;

(3) $y= 0.5$, that is, the expansion is along the 
disorder line where the ground states are known exactly.

For the cases of $y=0$ and $0.2411$, 
the series have been computed to order $\lambda^{23}$ for the ground 
state energy $E_0$, to order $\lambda^{13}$ for antiferromagnetic susceptibility
$\chi$, and to order $\lambda^{11}$ for the triplet dispersion.
There are only 12 graphs that contribute to the ground state energy and
dispersion, and 14 graphs that contribute to the 
antiferromagnetic susceptibility. This considerably extends
previous series expansions for this model \cite{sg89,bar98}.

For the case of $y=0.5$, the series
are carried out to order $\lambda^{23}$ for the dispersion and
to order $\lambda^{17}$ for the susceptibility and the spectral
weight. Due to some special symmetries of the model along the disorder
line, a graph with $n$ dimers contributes first
in order $2(n-1)$. Thus only 8 graphs are needed to carry out the
expansions complete to order $\lambda^{17}$ and only 12 graphs to
carry them out to order $\lambda^{23}$. For this model, the dispersion
is symmetric around $q=\pi/2$, whereas the spectral weight 
at $q$, $W(q)$, is related to that at $\pi-q$ by the relation:
\begin{equation}
W(q)(1-\cos{(\pi-q)})=W(\pi-q)(1-\cos{q}).
\end{equation}
It is known that at $q=\pi/2$
the triplet dispersion and its spectral weight
do not change with $\lambda$ \cite{J1-J2-delta}.
In the perturbation expansion this result is reflected in the
fact that the expansion coefficients after the zeroeth order vanish.
This serves as a further check on the calculations.
The expansion coefficients would be available on request.

\section{The log corrections in the power-law correlated phase}
For the case of $y< y_c$, the asymptotic
behavior for ground state energy $E_0$, the energy gap $\Delta$ and
antiferromagnetic susceptibility $\chi$ as $\delta \to 0$ ($\lambda \to 1$) 
are known to be\cite{aff89}:
\begin{eqnarray}
E_0(\delta ) - E_0 (\delta =0) && \propto 
{ \delta^{4/3} \over \vert \ln \delta/\delta_0 \vert^a } \nonumber \\ 
\Delta (\delta )  && \propto 
{\delta^{2/3} \over \vert \ln \delta/\delta_0 \vert^b }  \label{eqlog} \\
\chi (\delta )  && \propto 
\delta^{-2/3}  \vert \ln \delta/\delta_0 \vert^c  \nonumber 
\end{eqnarray}
with $a=1$ and $b=1/2$\cite{aff89}, $c$ has not been computed previously, as far as we are aware.
Here, the logarithmic corrections are due to the marginal operators
present in the model.
It has been argued that these marginal operators
are absent at the transition $y=y_c$ to the dimerized phase 
and we expect to have pure power-law asymptotic behavior there.

To study their behavior, the series were analysed using the 
standard Dlog Pad\'e approximants. These approximants completely miss
possible logarithmic corrections and thus can only lead to
``effective" power-law exponents. 
The estimates for the critical points and exponents from the $[n/m]$
Dlog Pad\'e approximants to the series for energy gap and antiferromagnetic
susceptibility are given in Table I. From this table,
we see that the critical point lies at $\lambda_c =1.00(1)$ as expected.
The ``effective" critical exponents based on unbiased estimates  (UB)
and estimates with critical point biassed at $\lambda_c=1$ (B)
are given in Table II. We can see that for the case of $y=y_c$, 
the exponents agree with $\nu=\gamma=2/3$ very well,
This provides support to the argument that
logarithmic corrections are absent here.
For $y=0$, the ``effective" critical exponents for both
$\Delta$ and $\chi$ are quite different from $2/3$. As argued by Affleck
and Bonner \cite{affleck-bonner}, the logarithmic corrections lead
to ``effective" exponents which vary slowly with the size of the system,
or the length of the series. The estimated exponent values are in between
the true values and the effective exponents for size 20 calculated
by them. One could also attempt to directly study the logarithmic
singularity by multiplying the series by an appropriate power of
$|\log{\delta/\delta_0}|$ before carrying out the Dlog Pad\'e
analysis. However, such an analysis will depend on the choice of $\delta_0$.
Such an analysis, varying $\delta_0$ will not be attempted here.
We note simply that choosing $\delta_0=1$, moves the effective exponents
too far in the opposite direction.

\section{Dispersion and Spectral Weight along the Disorder Line}

In this section, we begin by calculating the triplet dispersion
as a function of $\delta$.
The dispersion-relations are shown for
a number of $\delta$ values in figure 1.
As one approaches $\delta=0$, the gap in the spectrum stays
robust: it approaches a constant 
with correction proportional to $(1-\lambda)^{2/3}$\cite{sushkov}, 
so in series extrapolation we transform the series to a new variable
\begin{equation}
\lambda' = 1- (1-\lambda)^{2/3}~,
\end{equation}
to remove the singularity at $\lambda=1$. 
For $\delta=0$, the spectrum compares well with the
lowest lying triplet-states in the
variational calculation of Sutherland and Shastry \cite{shastry}.
As remarked earlier, at $q=\pi/2$ the triplet state remains unchanged
as a function of $\delta$.

The spectral weight of the triplets undergoes dramatic
changes as the dynamical phase transition is approached.\cite{sushkov} 
Over substantial portions of the Brillouin zone, the spectral weight
vanishes as $\delta\to 0$. A simple Dlog Pad\'e analysis of the
spectral weight series gives a vanishing spectral weight at
$\lambda$ slightly less than unity
($\delta$ slightly larger than zero), with an exponent which varies
with the estimated critical point. It is difficult to determine
this exponent accurately in an unbiassed manner. Since it is
known that the spectral weights vanish as $\delta^{1/3}$ 
we adopt the following series extrapolation scheme: For a given wavevector,
we generate the series in $\lambda$ 
for the spectral-weight divided by $(1-\lambda)^{1/3}$.
For a range of wavevectors the Pad\'e approximant for the resulting series
converges very well. For wavevectors close to $\pi/2$, the resulting
series diverges as $\lambda\to 1$. This shows that for these $q$-values
the spectral weight remains finite and is thus analyzed by a
direct analysis of the spectral weight series [without the division
by $(1-\lambda)^{1/3}$]. The resulting spectral weight at a few values
of $\delta$ are shown in figure 2.

The susceptibility remains finite as $\delta\to 0$.
Rather than look for a weak singularity in the susceptibility
as $\delta\to 0$, we analyze the singular structure of the susceptibility
at imaginary wavevectors. We expect that for $\delta\ne 0$, the
susceptibility for small $k=\pi-q$ should have the form,
\begin{equation}
\chi(k)\approx {A\over 1+k^2\xi^2}
\end{equation}
So that at imaginary wavevector $k=i/\xi$, the susceptibility
has a simple pole. However, as $\delta\to 0$, the spectrum now
consists of two-spinon continuum, and the static susceptibility
should now have a branch cut of the form \cite{spinon2},
\begin{equation}
\chi(k)\approx {A\over (1+k^2\xi^2)^{\alpha}}, \label{eq10}
\end{equation}
with an exponent $\alpha<1$.
Since the correlation length varies smoothly as a function of
$\lambda$, this implies that if we consider the series for the
susceptibility at a fixed imaginary wavevector, $\kappa$, it should have
a singularity at the $\lambda$ value
where the correlation length becomes $i/\kappa$. This singularity should
be a simple pole ( exponent unity) which should reduce to a branch
cut ( exponent less than unity) as
$\lambda\to 1$ ($\delta\to 0$). 
We calculated the series for the susceptibility at
a number of imaginary wavevectors $\kappa$, which were then analyzed by Pad\'e
approximants. The location of the singularity tells us the
$\delta$ value at which the correlation length $\xi$  equals $i/\kappa$.
Thus this analysis gives both the correlation length and the exponent
$\alpha$ as a function of our parameter $\delta$.
The resulting exponents and correlation length
are plotted as a function
of $\delta$ in Figure 3. The change in the nature of the
singularity is clearly evident from the plot.

\section{Conclusions}
In this paper we have used series expansion methods to study the
spin-half Heisenberg chain with bond alternation and nearest
and second neighbor interactions. Our results are consistent
with previous ones which show that ``effective" exponents
are modified due to logarithmic corrections in the power-law
correlated phase of this model,
but these modifications go away when the system is tuned to
the critical point separating the power-law and spontaneously
dimerized phases. We have also studied in detail the
triplet spectra along the disorder line, where the ground
states are known exactly. Our results provide clear evidence
for a dynamical transition from triplet elementary excitations to
spinons. The vanishing of the spectral weight and the change in the
singularity structure of the wavevector dependent static susceptibility
exhibit such a transition, while the ground state
(and hence all equal-time correlation functions)
remains free of singularities.
This method should prove useful in looking for
spin-half excitations in spin systems for $d>1$.

\acknowledgments
We would like to thank Valeri Kotov and Oleg Sushkov for many
useful discussions. This work is supported in part by a grant
from the National Science Foundation (DMR-9616574) (R.R.P.S),
and by the Australian Research Council (Z.W.).
Part of the computation in this work has been done using the 
facilities of the New South Wales 
Centre for Parallel Computing.

% \newpage

%=======================================================================
\begin{figure}[h] %h: here; t:top of page; b:bottom of page; p: page of float
\vspace{-1cm}
%\par
\centerline{\hbox{\psfig{figure=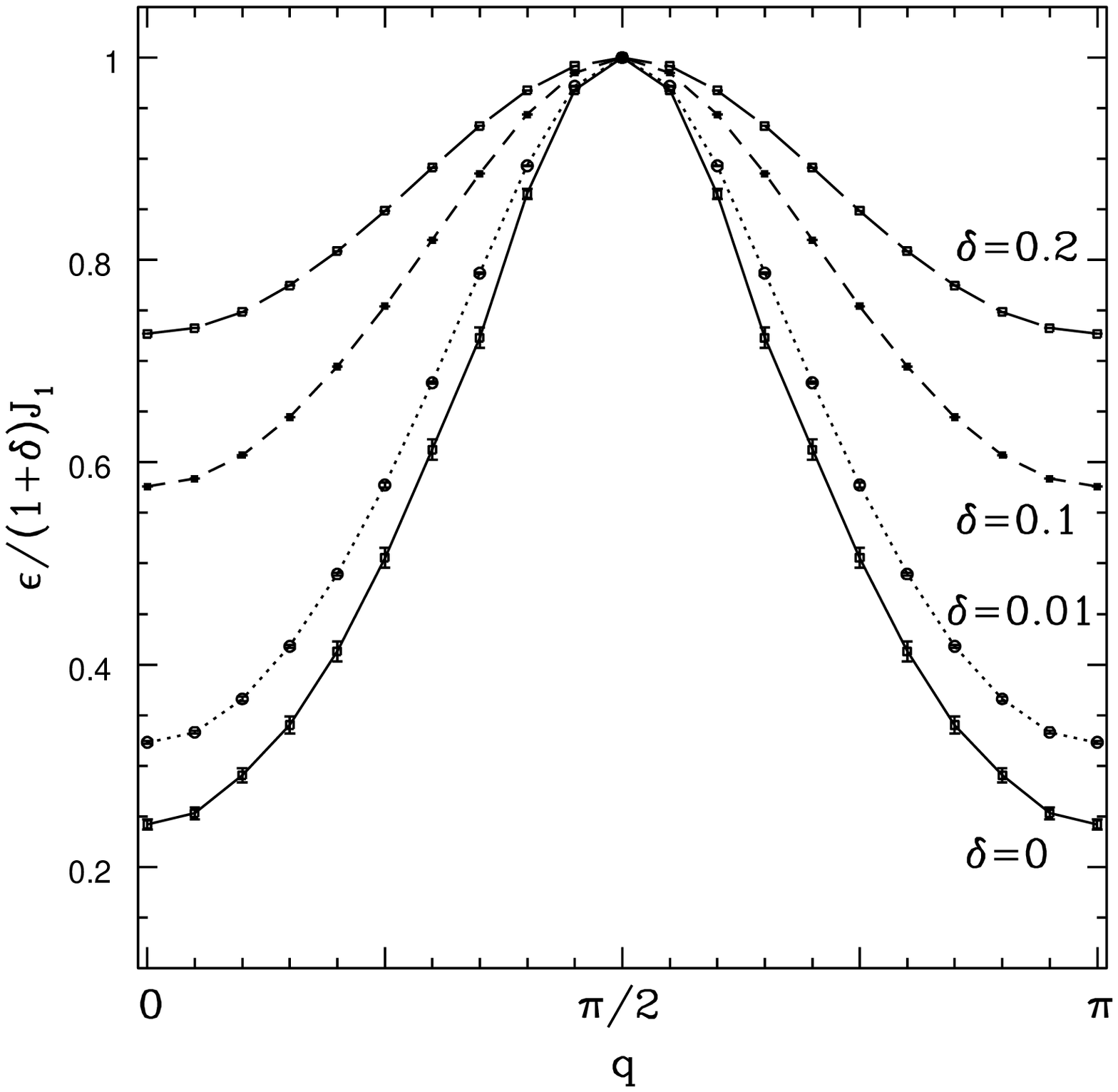,width=9cm}}}
%\par
\vspace{-2cm}
\caption{The excitation  spectrum
$\epsilon(q) /(1+\delta)J_1$ for 
the $J_1-J_2-\delta$ spin-half chain along the
disorder line, for $\delta=0,0.01,0.1,0.2$.
}
\label{fig1}
\end{figure}
%=======================================================================

%=======================================================================
\begin{figure}[h] %h: here; t:top of page; b:bottom of page; p: page of float
\vspace{-1.5cm}
%\par
\centerline{\hbox{\psfig{figure=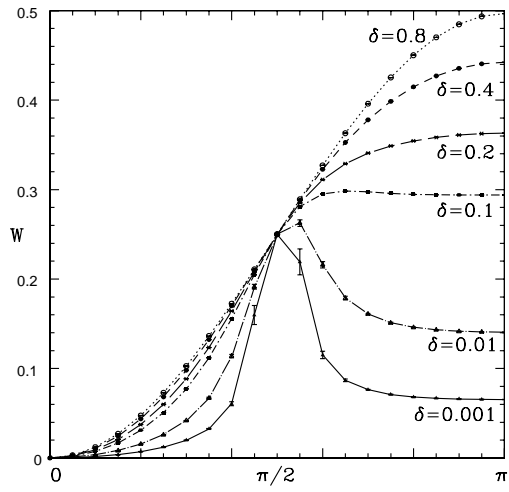,width=9cm}}}
%\par
\vspace{-2cm}
\caption{The spetral weight $W(q)$ for the $J_1-J_2-\delta$
spin-half chain along the disorder line
for $\delta=0.001,0.01,0.1,0.2,0.4,0.8$.
}
\label{fig2}
\end{figure}
%=======================================================================

%=======================================================================
\begin{figure}[h] %h: here; t:top of page; b:bottom of page; p: page of float
\vspace{-1.5cm}
%\par
\centerline{\hbox{\psfig{figure=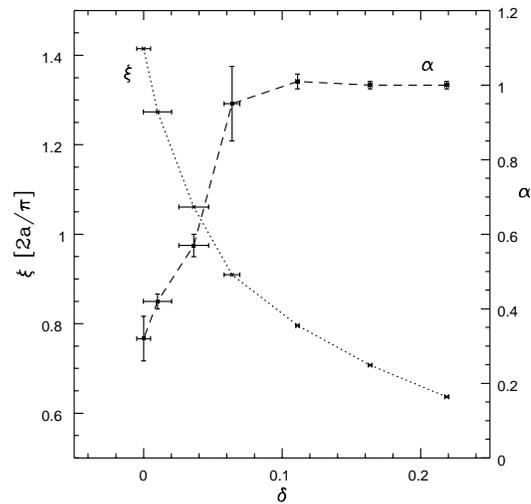,width=9cm}}}
%\par
\vspace{-2cm}
\caption{The correlation length $\xi$ and 
the critical exponent $\alpha$, representing the singularity in the
static susceptibility at imaginary wavevectors (Eq. \ref{eq10})
as a function of  $\delta$. Note that $\alpha=1$ represents a simple 
pole and implies that the elementary excitations are triplets,
whereas a smaller $\alpha$ represents a branch cut,
and implies that the triplets have become
composite objects.
}
\label{fig3}
\end{figure}
%=======================================================================

\begin{table}
\squeezetable
\setdec 0.00000000000
\caption{$[n/m]$ Dlog Pad\'e approximants to the series for energy gap $\Delta$ 
and antiferromagnetic susceptibility $\chi$. An asterisk denotes a defective
approximant.
}
 \label{tab1}
\begin{tabular}{rrrrrr}
\multicolumn{1}{c}{n} &\multicolumn{1}{c}{$[(n-2)/n]$}&\multicolumn{1}{c}{$[(n-1)/n]$}
&\multicolumn{1}{c}{$[n/n]$} &\multicolumn{1}{c}{$[(n+1)/n]$}&\multicolumn{1}{c}{$[(n+2)/n]$}
 \\
\multicolumn{1}{c}{} &\multicolumn{1}{c}{pole(residue)} &\multicolumn{1}{c}{pole(residue)} 
&\multicolumn{1}{c}{pole(residue)} &\multicolumn{1}{c}{pole(residue)} &\multicolumn{1}{c}{pole(residue)} \\
\tableline
\multicolumn{6}{c}{$\chi$ for $y=0$} \\
 n= 2 &                      & \dec 0.9759($-$0.749)  & \dec 1.1109($-$0.218)$^*$ & \dec 1.0143($-$0.864) & \dec 0.9950($-$0.771) \\
 n= 3 & \dec 0.9912($-$0.786) & \dec 1.0079($-$0.836)  & \dec 1.0045($-$0.822) & \dec 1.0033($-$0.816) & \dec 1.0030($-$0.814) \\
 n= 4 & \dec 1.0051($-$0.825) & \dec 1.0029($-$0.814)  & \dec 1.0029($-$0.813) & \dec 1.0040($-$0.818)$^*$ & \dec 0.9973($-$0.725)$^*$ \\
 n= 5 & \dec 1.0029($-$0.813) & \dec 1.0029($-$0.814)$^*$  & \dec 0.9905($-$0.532)$^*$ & \dec 1.0021($-$0.808)0 & \dec 1.0012($-$0.798) \\
 n= 6 & \dec 0.9982($-$0.746)$^*$ & \dec 1.0012($-$0.798)  & \dec 1.0010($-$0.797) & \\
\tableline
\multicolumn{6}{c}{$\Delta$ for $y=0$}\\
 n= 2 &                       & \dec 1.1088(0.970) & \dec 0.8216(0.322) & \dec 0.8374(0.348) & \dec 1.0986(1.243) \\
 n= 3 & \dec 0.9531(0.621) & \dec 0.9906(0.711) & \dec 1.0158(0.793) & \dec 0.9982(0.724) & \dec 1.0022(0.742) \\
 n= 4 & \dec 1.0495(0.986) & \dec 1.0047(0.751) & \dec 1.0016(0.738) & \dec 1.0018(0.740) & \dec 0.9896(0.501)$^*$ \\
 n= 5 & \dec 1.0021(0.741) & \dec 1.0018(0.739) & \dec 1.0017(0.739)  \\
\tableline
\multicolumn{6}{c}{$\chi$ for $y=0.2411$} \\
 n= 3 & \dec 0.9931($-$0.653) & \dec 1.0075($-$0.712) & \dec 1.0002($-$0.676) & \dec 1.0006($-$0.678) & \dec 1.0005($-$0.677) \\
 n= 4 & \dec 1.0004($-$0.677) & \dec 1.0005($-$0.677) & \dec 1.0005($-$0.677) & \dec 1.0008($-$0.678)$^*$ & \dec 1.0002($-$0.674) \\
 n= 5 & \dec 1.0008($-$0.680) & \dec 1.0005($-$0.677) & \dec 1.0002($-$0.674) & \dec 1.0001($-$0.673) & \dec 1.0001($-$0.673) \\
 n= 6 & \dec 1.0005($-$0.677) & \dec 1.0001($-$0.673) & \dec 1.0001($-$0.673) &   \\
\tableline                 
\multicolumn{6}{c}{$\Delta$ for $y=0.2411$} \\
 n= 2 &                       & \dec 1.6581(3.412) & \dec 1.4233(2.031) & \dec 0.6121(0.043) & \dec 0.7576(0.143) \\
 n= 3 & \dec 1.0588(0.737) & \dec 0.7751(0.190)$^*$ & \dec 1.1039(1.021) & \dec 0.9923(0.620) & \dec 1.0011(0.653) \\
 n= 4 & \dec 0.9562(0.531) & \dec 1.0082(0.680) & \dec 1.0002(0.649) & \dec 0.9960(0.632) & \dec 0.9945(0.626) \\
 n= 5 & \dec 1.0017(0.656) & \dec 1.0512(0.670)$^*$ & \dec 0.9944(0.626)$^*$ &  \\
\end{tabular}                                                        
\end{table}

\begin{table}
\squeezetable
\setdec 0.00000000000
\caption{Estimates of ``effective" critical exponents obtained 
by Dlog Pad\'e approximants
to the series for susceptibility $\chi$, the energy gap $\Delta$,
and the difference of the ground state energy 
$E_0(\lambda ) - E_0 (\lambda =1)$.
% and the estimates of critical exponents  
% to the series with pure power-law
% exponents by dividing out the expected lagarithmic corrections.
Both unbiased estimates  (UB)
and estimates biased critical point $\lambda_c=1$ (B) are listed.
}
 \label{tab2}
\begin{tabular}{r|rr|rr}
\multicolumn{1}{c|}{series} &\multicolumn{2}{c|}{$y=0$}
&\multicolumn{2}{c}{$y=0.2411$} \\
\multicolumn{1}{c|}{} &\multicolumn{1}{c}{UB}&\multicolumn{1}{c|}{B}
&\multicolumn{1}{c}{UB}&\multicolumn{1}{c}{B} \\
\tableline
$\Delta$                    & $\nu = 0.74(3)$ & $\nu = 0.72(3)$ & $\nu = 0.65(3)$ & $\nu = 0.65(2)$ \\
$\chi$                      & $\gamma = 0.80(3)$ & $\gamma = 0.78(2)$ & $\gamma = 0.675(10)$ & $\gamma = 0.675(8)$ \\
$E_0(\delta)-E_0(\delta=0)$ & $\alpha=0.95(4)$ & $\alpha = 0.97(2)$ \\
\end{tabular}
\end{table}

\end{document}